\documentclass[superscriptaddress,twocolumn,amsmath,amssymb,nofootinbib,longbibliography]{revtex4-1}

\usepackage{graphicx}

\usepackage[usenames,dvipsnames]{xcolor}

\newcommand{\tens}[1]{{\boldsymbol{#1}}}       
\newcommand{\ts}[1]{{\boldsymbol{#1}}}         

\newcommand{\grad}{{\tens{d}}}                 
\newcommand{\cv}[1]{{\tens{\partial}}_{#1}}    
\newcommand{\pa}{\partial}                     

\newcommand{\A}[1]{A^{\!(#1)}}                 

\newcommand{\dg}{{N}}                            

\newcommand{\EMA}{\mathsf{A}}                     
\newcommand{\EMF}{\mathsf{F}}                     

\newcommand{\scs}{C}

\newcommand{\be}{\begin{equation}}             
\newcommand{\ee}{\end{equation}}               

\begin{document}

\title{Massive Vector Fields in Rotating Black-Hole Spacetimes: Separability and Quasinormal Modes}

\author{Valeri P. Frolov}
\email{vfrolov@ualberta.ca}
\affiliation{Theoretical Physics Institute, University of Alberta, Edmonton,
Alberta, Canada T6G 2E1}
\author{Pavel Krtou\v{s}}
\email{Pavel.Krtous@utf.mff.cuni.cz}
\affiliation{Institute of Theoretical Physics, Faculty of Mathematics and Physics, Charles University,\\
V~Hole\v{s}ovi\v{c}k\'ach~2, Prague, 18000, Czech Republic}
\author{David Kubiz\v n\'ak}
\email{dkubiznak@perimeterinstitute.ca}
\affiliation{Perimeter Institute, 31 Caroline Street North, Waterloo, ON, N2L 2Y5, Canada}
\author{Jorge E. Santos}
\email{jss55@cam.ac.uk}
\affiliation{DAMTP, Centre for Mathematical Sciences, Wilberforce Road, Cambridge, CB3 0WA, United Kingdom}

\date{April 1, 2018}            

\begin{abstract}
We demonstrate the separability of the massive vector (Proca) field equation in general \mbox{Kerr--NUT--(A)dS} black hole spacetimes in any number of dimensions, filling a long-standing gap in the literature. The obtained separated equations are studied in more detail for the four-dimensional Kerr geometry  and the corresponding quasinormal modes are calculated. Two of the three independent polarizations of the Proca field are shown to emerge from the separation ansatz and the results are found in an excellent agreement with those of the recent numerical study where the full coupled partial differential equations were tackled without using the separability property.
\end{abstract}

\maketitle
\section{Introduction}
\label{sc:intro}

Maxwell equations describe a theory of a massless vector field (photons). When the massive term is added, a theory of the massive spin-1 vector particle is obtained. The massive term breaks the gauge invariance. Historically, the massive vector mesons were first used by Proca \cite{Proca:1936} in his attempt to explain the short-range nuclear forces
(see also \cite{Belinfante:1949,Rosen:1994}). In modern physics the Proca equation is used in the standard model for describing the massive spin-1 $Z$ and $W$ bosons. The Proca equation, originally formulated in a flat spacetime, can easily be generalized to the curved background, see, e.g., \cite{Seitz:1986sc}. Recently, the interest in the Proca fields greatly increased after it was demonstrated that ultralight vector fields might be responsible for instability of rotating black holes.

The ultralight massive vector fields, such as dark photons, are a feature of many beyond standard model physics scenarios, string theory for example \cite{Goodsell:2009xc}. Together with axion-like scalars such fields provide a compelling candidate for cold dark matter \cite{Essig:2013lka}. However, due to their tiny mass and weak coupling, direct searches for such particles are challenging, and alternative methods for detection are sought. One such promising method is to look for the superradiant instabilities of rotating black holes, leading to a formation of the bosonic condensate outside the horizon, and a subsequent gravitational wave emission and spin-down of the  black hole. Consequently, gravitational and electromagnetic wave observations of astrophysical black holes provide direct constraints on certain models of ultralight massive particles, e.g., \cite{Arvanitaki:2009fg, Baryakhtar:2017ngi, Cardoso:2018tly}.

In order to make these predictions, one has to (in the first approximation) study the behavior of the corresponding test fields in the background of the rotating black hole. For the massive vector field this is a formidable task, as the corresponding equations were believed not to separate \cite{Pani:2012vp, Rosa:2011my, Pani:2012bp} and the problem was studied either using certain approximations, e.g.~\cite{Pani:2012vp,Pani:2012bp,Endlich:2016jgc,Baryakhtar:2017ngi}, or by applying heavy numerics \cite{East:2017ovw, East:2017mrj, Cardoso:2018tly}.

In this Letter, we show that no such limitations are necessary. Namely, we demonstrate that with a proper ansatz the massive vector field equation does  separate and that this not only happens for the four-dimensional vacuum Kerr spacetime, but remains also true in higher dimensions and in the presence of the cosmological constant and NUT charges. Not only this result fills a long-standing gap in the literature on separability of physical field equations in the black hole spacetimes, e.g.~\cite{FrolovKrtousKubiznak:2017review}, but it also provides an opportunity to study the properties of these fields (for example their superradiant instability) in a significantly simplified manner.

In this Letter we proceed as follows. We first demonstrate separability of the massive vector field equations in the Kerr--NUT--(A)dS black hole spacetimes  in any number of dimensions \cite{Chen:2006xh} and relate this separability to the existence of the hidden symmetry present in these spacetimes \cite{FrolovKrtousKubiznak:2017review}. This result is new and non-trivial even in four dimensions---for the Kerr geometry. Next, restricting to this case, we demonstrate that our ansatz correctly captures two of the  three independent polarizations of the Proca field and that we are able to much more effectively reconstruct the unstable modes recently studied in \cite{Cardoso:2018tly} by heavy numerical methods. We conclude with some general remarks on possible future directions.

\section{Spacetime geometry}
\label{sc:geometry}

In what follows we want to demonstrate the separability of the four- and higher-dimensional Proca equations in the background of a wide class of metrics that include the general Kerr--NUT--(A)dS solutions \cite{Chen:2006xh} as a special case. Such a separability is valid in both even and odd dimensions, however, the forms of the metric and of the separated equations are slightly different. For brevity, in this Letter we restrict to the case of even dimensions, $D=2\dg$.

The metrics we consider are of the form
\be\label{metric}
\ts{g}
  =\sum_{\mu=1}^\dg\;\biggl[\; \frac{U_\mu}{X_\mu}\,{\grad x_{\mu}^{2}}
  +\, \frac{X_\mu}{U_\mu}\,\Bigl(\,\sum_{j=0}^{\dg-1} \A{j}_{\mu}\grad\psi_j \Bigr)^{\!2}
  \;\biggr]\, .
\end{equation}
Here, $\A{j}_{\mu}$ and $U_\mu$ are polynomials in $x_{\mu}^2$ defined by the following relations:
\begin{gather}
A_{\mu}(\beta)\equiv\prod_{\substack{\nu=1\\\nu\neq\mu}}^\dg (1+\beta^2 x_\nu^2)
   =\sum_{j=0}^{\dg-1}\A{j}_\mu\,\beta^{2j}
    \;,\\
U_{\mu}=\prod_{\substack{\nu=1\\\nu\ne\mu}}^\dg (x_{\nu}^2-x_{\mu}^2)\, ,
\end{gather}
and $X_\mu$ are arbitrary functions of one variable, $X_\mu(x_{\mu})$, cf.\ \cite{FrolovKrtousKubiznak:2017review}. We call the metric \eqref{metric} an {\em off-shell metric}, to distinguish it from a special case when \eqref{metric} obeys the Einstein equations with or without the cosmological constant. In such a case functions $X_{\mu}$ become special polynomials and the metric describes the Kerr--NUT--(A)dS solution \cite{Chen:2006xh}.

The characteristic property of the metric \eqref{metric} is that it admits the \emph{principal tensor $\ts{h}$}, see \cite{FrolovKrtousKubiznak:2017review}.  The principal tensor is a non-degenerate closed conformal Killing--Yano 2-form. It satisfies the equation
\begin{equation}
    \nabla_{\!c} h_{ab} = g_{ca}\xi_b - g_{cb}\xi_a\;,\label{PrincTdef}
\end{equation}
where $\ts{\xi}$ is a primary Killing vector,
\begin{equation}\label{PrimaryVec}
    \xi_a = \frac1{D-1}\nabla^b h_{ba}\;.
\end{equation}

The off-shell metric possesses $\dg$ Killing vectors
\be
\ts{l}_{(j)}=\cv{\psi_j}\, ,\quad j=0,\ldots \dg-1\, ,
\ee
the first of which coincides with the primary Killing vector, $\ts{l}_{(0)}=\ts{\xi}$. The remaining $\dg$ canonical coordinates, $x_{\mu}$,\ \ $\mu=1,\ldots,\dg$, are nothing but the eigenvalues of the principal tensor $\ts{h}$, which in the canonical coordinates $(x_{\mu},\psi_j)$ takes the following form:
\be
 \ts{h} = \sum_{\mu=1}^\dg x_\mu \, \grad x_{\mu}
   \wedge \left[\sum_{j=0}^{\dg-1}\A{j}_{\mu}\grad\psi_j\right]\,.
\ee

\section{Separation of variables}
\label{sc:separability}

Our goal is to demonstrate the separability of the Proca equation
\be\label{Proca}
\nabla_{\!n}\EMF^{an} +m^2 \EMA^a=0\,
\ee
in the metric \eqref{metric}. Here $\EMF_{an}=\nabla_a \EMA_n-\nabla_n \EMA_a$, and $m$ is the mass of the massive vector field $\ts{\EMA}$. To this purpose we shall use the following ansatz \cite{Lunin:2017,FrolovKrtousKubiznak:2018a,KrtousEtal:2018}:
\be\label{ans}
  \EMA^a = B^{ab} \nabla_{\!b}Z\, ,\quad
  B^{ab} (g_{bc}+i\mu h_{bc}) = \delta^a_c\,.
\ee
Here, $\mu$ is a complex parameter, and the potential function $Z$ is written in the multiplicative separated form
\be\label{multsep}
Z=\prod_{\nu=1}^{\dg}R_\nu(x_{\nu})\exp{\Bigl(i\sum_{j=0}^{\dg-1}L_j\psi_j\Bigr)}\, .
\ee

An immediate consequence of the Proca equation is the ``Lorenz condition'',
\be\label{Lor}
\nabla_{\!a}\EMA^a=0\, .
\ee
As shown in \cite{KrtousEtal:2018}, for the ansatz \eqref{ans}, this explicitly reads
\be
\nabla_{\!a}\EMA^a = \frac{Z}{A}\,\sum_{\nu=1}^{\dg} \frac{A_\nu}{U_\nu}\,
    \frac{1}{R_\nu}{\cal D}_\nu {\cal R_\nu}\, ,\label{AA}
\ee
where
\be
\begin{split}
\mathcal{D}_\nu&=q_{\nu}\frac{\pa}{\pa x_\nu}\biggl[\frac{X_{\nu}}{q_{\nu}}\frac{\pa}{\pa x_\nu}\biggr]
  - \frac1{X_\nu} \Bigl[\sum_{j=0}^{\dg-1} (-x_\nu^2)^{\dg{-}1{-}j}L_j\Bigr]^2\\
&\quad+ \mu\frac{2-q_{\nu}}{q_{\nu}}(-\mu^2)^{(1-\dg)}\sum_{j=0}^{\dg-1}(-\mu^2)^{j}L_j\, ,
\end{split}\raisetag{8ex}
\ee
and
\be
q_{\nu}=1-\mu^2 x_{\nu}^2\,,\quad A=\prod_{\nu=1}^\dg q_\nu\, .
\ee
The Lorenz condition \eqref{Lor} may be satisfied provided the mode functions $R_{\nu}$ obey the separated equations
\be\label{DD}
{\cal D}_\nu R_{\nu}=\tilde{\scs}_{\nu} R_{\nu}\,,
\ee
where $\tilde{\scs}_{\nu}$ are polynomials in variable $x_\nu^2$ with the same coefficients $\scs_k$,
\be
\tilde{\scs}_{\nu} = \sum_{k=0}^{\dg-1} \scs_k\, (-x_\nu^2)^{\dg{-}1{-}k}\, .
\ee
Indeed, in such a case the expression in \eqref{AA} sums to
\begin{equation}\label{AAA}
    \nabla_{\!a}\EMA^a = \frac{Z}{A}\, \sum_{j=0}^{\dg-1} \scs_j\, (-\mu^2)^{j}\;,
\end{equation}
and we see that the Lorenz condition holds provided the parameter~$\mu$ satisfies the following constraint:
\begin{equation}\label{C=0}
    \sum_{j=0}^{\dg-1} \scs_j\, (-\mu^2)^{j} = 0\;.
\end{equation}

The results of \cite{KrtousEtal:2018} can be also used to find the representation of $\nabla_{\!n}\EMF^{an}$ for the ansatz \eqref{ans}. When {the Lorenz condition \eqref{Lor} holds,} one has
\be
{\nabla_{\!n}\EMF^{an} +m^2 \EMA^a= - B^{am}\nabla_{\!m} J\, ,}
\ee
with
\be
{J=\Box Z+2\beta \xi_k B^{kn}\nabla_{\!n}Z - m^2 Z \, .}
\ee
The key observation is that $J$ also separates in the form
\be{
J= Z \sum_{\nu=1}^\dg \frac{1}{U_\nu}\frac1{R_\nu}
   \bigl[\mathcal{D}_{\nu}-m^2 (-x_{\nu}^2)^{\dg-1}\bigr]R_\nu\, ,}
\ee
where the mass term has been rewritten using the identity:
\be
\sum_{\nu=1}^\dg \frac{1}{U_{\nu}}\,(-x_{\nu}^2)^{\dg-1-j}=\delta^j_0
\ee
for $j=0$. The same identity guarantees ${J=0}$, provided the modes $R_{\nu}(x_{\nu})$ obey separated equations \eqref{DD} with an additional condition that the coefficient $\scs_0$ of the highest order term in the polynomials~$\tilde{\scs_{\nu}}$ is given by the mass,
\begin{equation}\label{C0m2}
    {\scs_0=m^2\;.}
\end{equation}
This finishes the proof of separability of the Proca equation in the off-shell Kerr--NUT--(A)dS metrics.

Summarizing, the Proca equation \eqref{Proca} for the vector field $\ts{\EMA}$ in the form \eqref{ans} can be solved using the multiplicative separation ansatz  \eqref{multsep}, where the mode functions $R_\nu$ satisfy ordinary differential equations (ODEs) \eqref{DD} with separation constants $\scs_k$ and $L_j$, provided that $\scs_0$ is given by \eqref{C0m2} and $\mu$ satisfies \eqref{C=0}. The obtained mode functions are thus labeled by the full set of $2\dg-1=D-1$ separation constants, and in that sense, they are completely general. However, at the moment it is not clear how the polarizations of the field are captured by our ansatz. One can speculate that the choice of the root $\mu$ in the constraint \eqref{C=0} could be related to the choice of polarization. If so, the obtained solutions describe $N-1$ complex polarization modes, that is, $D-2$ real modes, and only one polarization is missing. This speculation is supported by the numerics described below. However, further discussion of this issue is necessary.

\section{Quasi-normal modes}
\label{sc:QNM}
The demonstrated separability of the Proca equation has many interesting applications. For example, it allows one to study the problem of the stability of rotating black holes with respect to the massive vector field condensation. This problem reduces to a study of the spectrum of separated ODEs and the corresponding calculation of the complex frequency of non-radiative quasinormal modes (sometimes called quasi-bound states or unstable modes). For original (non-separated) equations, the required numerical calculations are very time consuming, as one has to solve a system of partial differential equations (PDEs). The study of quasinormal modes by using the separated Proca equation has many advantages.

In order to illustrate this, we focus on the case of four-dimensional Kerr spacetime (${\dg=2}$) studied recently in \cite{Cardoso:2018tly}, setting both the NUT parameter and the cosmological constant equal to zero.
We will show that our separation ansatz correctly reproduces two of the three physical polarizations of the massive vector in this spacetime. The polarizations captured by our analysis are labeled by $S=\pm1$, and they reduce to those explored in \cite{Rosa:2011my} when the black hole spin vanishes (see also \cite{Konoplya:2006gq,Konoplya:2005hr}).

Before proceeding, let us translate our results to a more familiar Boyer--Lindquist frame. First, we change from the canonical coordinates $(\psi_0,x_1,x_2,\psi_1)$ to new coordinates $(t,r,\theta,\phi)$ via the map
\be
(\psi_0,x_1,x_2,\psi_1)=(t-a\,\phi,i\,r,a\,\cos\theta,\phi/a)\,,
\ee
and identify
\be
X_1 = 2M r-r^2-a^2\equiv -\Delta\,, \quad X_2 =-a^2\sin^2\theta\,.
\ee
Upon this, the metric element \eqref{metric} yields the standard Boyer--Lindquist form of the Kerr geometry. The black-hole horizon is a null hypersurface located at $r=r_+\equiv {M+\sqrt{M^2-a^2}}$; we are interested in the subextremal limit for which $M<|a|$.
Finally, we want to map the eigenvalues $L_0$ and $L_1$, to the eigenvalues of $i \partial_t$ and $-i\partial_\phi$, $\omega$ and $m_\phi$. This is accomplished via the linear map
\be
L_0 = -\omega\,,\qquad L_1 = a(m_\phi-a \omega)\,.
\ee
Note that since $\phi$ has period $2\pi$, we have $m_{\phi}\in\mathbb{Z}$.

Equations \eqref{DD} reduce to two differential equations in $r$ and $\theta$, respectively, which only couple to each other via their dependence on the parameters $\{\mu, \omega, m_\phi, m, M, a\}$:
\begin{subequations}
\begin{align}
&\frac{d}{d r}\!
  \left[\frac{\Delta}{q_r}\frac{d R}{d r}\right]
  +\left[\frac{K_r^2}{q_r \Delta}+\frac{2-q_r}{q_r^2}\frac{\sigma}{\mu}-\frac{m^2}{\mu^2}\right]R=0\,,
\\
&\frac{1}{\sin\theta}\frac{d}{d \theta}\!
   \left[\frac{\sin \theta}{q_\theta}\frac{d S}{d \theta}\right]
   -\left[\frac{K_\theta^2}{q_\theta \sin^2\theta}{+}\frac{2{-}q_\theta}{q_\theta^2}\frac{\sigma}{\mu}{-}\frac{m^2}{\mu^2}\right]S=0\,,
\end{align}
\label{eqs:coupled}
\end{subequations}
where
\be
\begin{gathered}
K_r = a\,m_{\phi}-(a^2{+}r^2)\omega\,,\quad K_\theta = m_{\phi}-a\,\omega\,\sin^2\theta\,,
\\
q_r = 1+\mu^2r^2\,,\quad q_{\theta} = 1-\mu^2a^2\cos^2\theta\,,
\\
\sigma = a \mu ^2 \left(m_{\phi }-a \omega \right)+\omega\,.
\end{gathered}
\ee
Note that at this stage, we have set
\be
\scs_0 = m^2\,,\quad
\scs_1={m^2}/{\mu^2}\,,
\ee
so the constraints \eqref{C=0} and \eqref{C0m2} are readily solved.
\begin{figure*}
\centering
\includegraphics[width=0.9\linewidth]{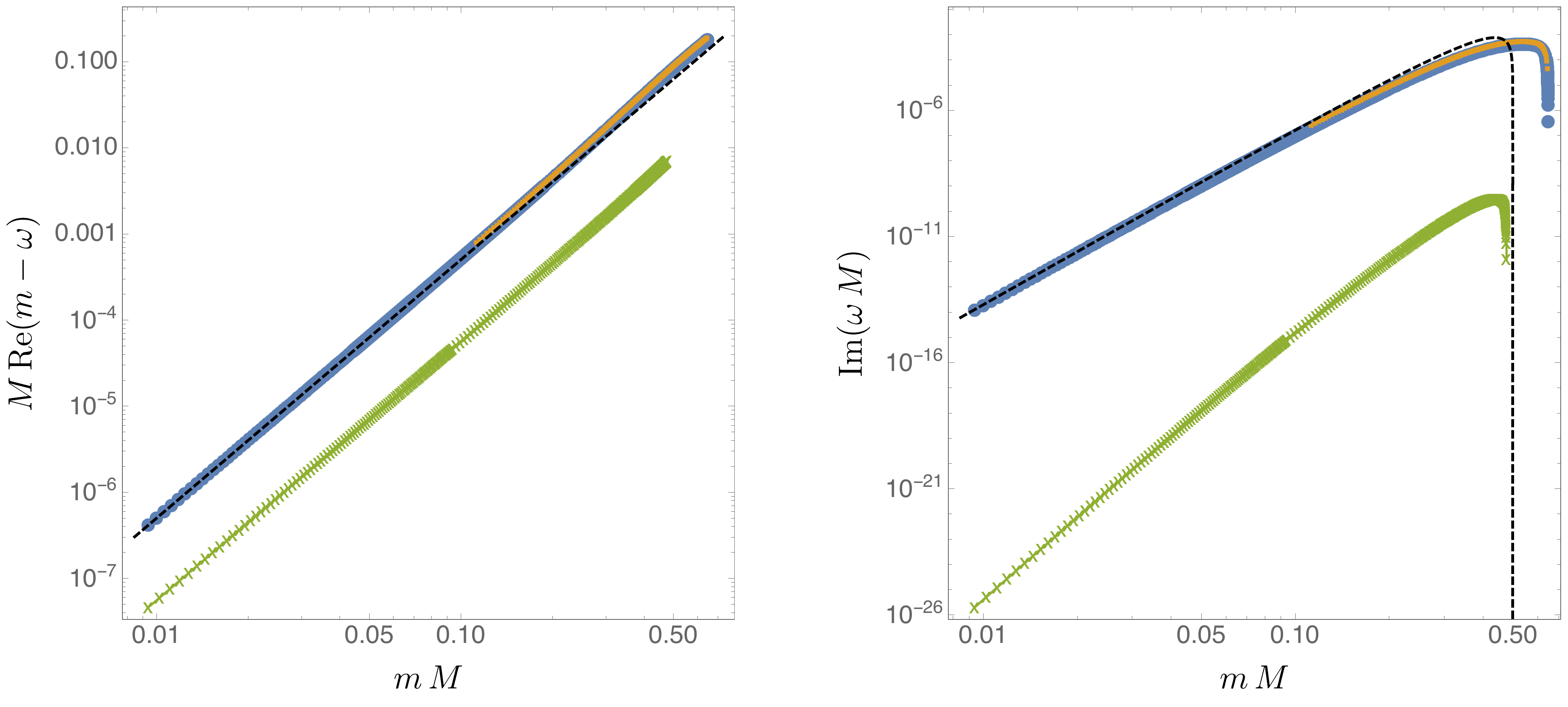}
\caption{\label{fig:graphs} {\bf Quasinormal modes.} {We display the frequency $\omega M$ as a function of $m M$ computed for quasinormal modes with $m_\phi=1$ and $a/M=0.998$. The real (left) and imaginary (right) part of $\omega M$ is displayed for the two polarizations $S=-1,+1$, with the top (online blue) curve corresponding to $S=-1$, and the lower (online green) one to $S=+1$. The polarization $S=-1$ corresponds to the most unstable mode. For this case, the (online orange) squares represent the data from \cite{Cardoso:2018tly}, and the dashed line gives the analytic approximation \eqref{eq:quasi} derived in \cite{Baryakhtar:2017ngi}. Solving separated ODEs allowed us to extend the numerics below $mM\sim10^{-2}$, compared to the range $mM\gtrsim10^{-1}$ of \cite{Cardoso:2018tly}. The agreement of the analytic approximation with our numerics for small $m M$ is excellent.}}\vspace{1ex}
\end{figure*}

Equations \eqref{eqs:coupled} are then solved as a coupled eigenvalue problem, with an eigenvalue pair $(\mu,\omega)$. The numerical method follows \emph{mutatis mutandis} \cite{Cardoso:2018tly}, and will not be detailed here again. The boundary conditions are chosen so that the modes are regular on the black-hole horizon and have finite energy. In Fig.~\ref{fig:graphs} we show the results of determining the lowest lying quasinormal mode for $a/M=0.998$, and $m_{\phi}=1$. The (blue) circles represent our numerical data, and the (orange) squares the numerical data of \cite{Cardoso:2018tly}. The values agree with each other to within $10^{-8}\%$.

The advantage of solving \eqref{eqs:coupled} instead of the coupled PDE system in \cite{Cardoso:2018tly} is immense. For instance, we were able to extend the polar mode of \cite{Cardoso:2018tly} with $S=-1$ and $m_{\phi}=1$, which is the most unstable mode, to masses as small as $m\,M\sim 10^{-3}$. In \cite{Baryakhtar:2017ngi}, the quasinormal mode spectrum of this mode was computed in the small mass regime $\widetilde{m}\equiv m\,M\ll1$, and it was found to be given by
\begin{equation}
M\omega \,\,\approx\,\, \widetilde{m}\left(1-\frac{\widetilde{m}^2}{2}\right)+2\,i\,\left(\frac{a}{M}-2\,\widetilde{m}\right)\,\widetilde{m}^7\,.
\label{eq:quasi}
\end{equation}
When $mM\ll1$, the PDEs of \cite{Cardoso:2018tly} become very hard to solve, so the regime of validity of this approximation was never identified. Here, we close this gap in the literature and compare our exact numerical results in Fig.~\ref{fig:graphs}, depicted as (blue) circles, to the analytic approximation \eqref{eq:quasi} represented as the dashed black line. When $m\,M\ll1$, the agreement is excellent. Additionally to modes calculated in \cite{Cardoso:2018tly}, we have also computed the lowest lying quasinormal mode with $S=+1$, which is depicted as the bottom (green) curve of Fig.~\ref{fig:graphs}. Our field ansatz thus covers two polarization modes---these reduce to the  even-parity (polar) scalar modes in the non-rotating limit. The scaling at small $m\,M$ of both polarizations is consistent with the one reported in \cite{Rosa:2011my,Pani:2012vp,Pani:2012bp,Baryakhtar:2017ngi}. The question of how to recover the last (axial) polarization remains currently open.

\section{Conclusions}
\label{sc:conclusion}

We have shown that the Proca field equation in a general Kerr--NUT--(A)dS spacetime separates in all dimensions. The key ingredient for this result is the ansatz (9). It is written purely in terms of the principal tensor, and thus, it encodes explicit and hidden symmetries of the spacetime.

Furthermore, we have used the separated equation to study the quasinormal mode spectrum of the Proca field around the Kerr black hole. Our spectrum matches that of \cite{Cardoso:2018tly} for two of the three polarizations. Except, solving just ODEs allows much more effective \mbox{calculations}.

The numerics confirm that our method captures more than one polarization. On general grounds, we have speculated that only one polarization mode is missing. However, the question as to how precisely these polarizations are described by the ansatz \eqref{ans} and how to obtain the remaining polarization(s) remains open.

Our results open a myriad of opportunities for the near future. At a more mathematical level, we could hope to use the results of \cite{Shlapentokh-Rothman:2013ysa} and prove the existence of exponentially growing modes for Proca perturbations around the Kerr black hole. Although both our work and \cite{Cardoso:2018tly} give ample numerical evidence that this is the case, it would be reassuring to have a definite mathematical proof. Also interesting, would be to repeat Detweiler's seminal calculation \cite{Detweiler:1980uk}, and then analytically extract the growth rate of Proca fields for $m M,a M\ll1$. Finally, we would like to mention that, still in four dimensions, we could study the stability of the Proca field for $\Lambda<0$, with the latter scenario being relevant in the context of the AdS/CFT correspondence. Perhaps more interestingly, virtually nothing is known about Proca perturbations in the context of higher-dimensional black holes. Our separation ansatz can be equally used in those circumstances.


\section*{Acknowledgements}
\label{sc:acknowledgements}

V.F.\ thanks the Natural Sciences and Engineering Research Council of Canada (NSERC) and the Killam Trust for their financial support. P.K.\ is supported by Czech Science Foundation Grant No.~\mbox{17-01625S}. D.K.\ acknowledges the Perimeter Institute for Theoretical Physics and the NSERC for their support. Research at Perimeter Institute is supported by the Government of Canada through the Department of Innovation, Science and Economic Development Canada and by the Province of Ontario through the Ministry of Research, Innovation and Science. J.E.S. is supported in part by STFC Grants No.~PHY-1504541 and No.~ST/P000681/1.


\begin{thebibliography}{24}%
\makeatletter
\providecommand \@ifxundefined [1]{%
 \@ifx{#1\undefined}
}%
\providecommand \@ifnum [1]{%
 \ifnum #1\expandafter \@firstoftwo
 \else \expandafter \@secondoftwo
 \fi
}%
\providecommand \@ifx [1]{%
 \ifx #1\expandafter \@firstoftwo
 \else \expandafter \@secondoftwo
 \fi
}%
\providecommand \natexlab [1]{#1}%
\providecommand \enquote  [1]{``#1''}%
\providecommand \bibnamefont  [1]{#1}%
\providecommand \bibfnamefont [1]{#1}%
\providecommand \citenamefont [1]{#1}%
\providecommand \href@noop [0]{\@secondoftwo}%
\providecommand \href [0]{\begingroup \@sanitize@url \@href}%
\providecommand \@href[1]{\@@startlink{#1}\@@href}%
\providecommand \@@href[1]{\endgroup#1\@@endlink}%
\providecommand \@sanitize@url [0]{\catcode `\\12\catcode `\$12\catcode
  `\&12\catcode `\#12\catcode `\^12\catcode `\_12\catcode `\%12\relax}%
\providecommand \@@startlink[1]{}%
\providecommand \@@endlink[0]{}%
\providecommand \url  [0]{\begingroup\@sanitize@url \@url }%
\providecommand \@url [1]{\endgroup\@href {#1}{\urlprefix }}%
\providecommand \urlprefix  [0]{URL }%
\providecommand \Eprint [0]{\href }%
\providecommand \doibase [0]{http://dx.doi.org/}%
\providecommand \selectlanguage [0]{\@gobble}%
\providecommand \bibinfo  [0]{\@secondoftwo}%
\providecommand \bibfield  [0]{\@secondoftwo}%
\providecommand \translation [1]{[#1]}%
\providecommand \BibitemOpen [0]{}%
\providecommand \bibitemStop [0]{}%
\providecommand \bibitemNoStop [0]{.\EOS\space}%
\providecommand \EOS [0]{\spacefactor3000\relax}%
\providecommand \BibitemShut  [1]{\csname bibitem#1\endcsname}%
\let\auto@bib@innerbib\@empty
\bibitem [{\citenamefont {Proca}(1936)}]{Proca:1936}%
  \BibitemOpen
  \bibfield  {author} {\bibinfo {author} {\bibfnamefont {A.}~\bibnamefont
  {Proca}},\ }\bibfield  {title} {\enquote {\bibinfo {title} {Sur la
  th\'{e}orie ondulatoire des \'{e}lectrons positifs et n\'{e}gatifs},}\
  }\href@noop {} {\bibfield  {journal} {\bibinfo  {journal} {Journal de
  Physique et le Radium}\ }\textbf {\bibinfo {volume} {7}},\ \bibinfo {pages}
  {347--353} (\bibinfo {year} {1936})}\BibitemShut {NoStop}%
\bibitem [{\citenamefont {Belinfante}(1949)}]{Belinfante:1949}%
  \BibitemOpen
  \bibfield  {author} {\bibinfo {author} {\bibfnamefont {Frederik~J.}\
  \bibnamefont {Belinfante}},\ }\bibfield  {title} {\enquote {\bibinfo {title}
  {The interaction representation of the {P}roca field},}\ }\href {\doibase
  10.1103/PhysRev.76.66} {\bibfield  {journal} {\bibinfo  {journal} {Phys.
  Rev.}\ }\textbf {\bibinfo {volume} {76}},\ \bibinfo {pages} {66--80}
  (\bibinfo {year} {1949})}\BibitemShut {NoStop}%
\bibitem [{\citenamefont {Rosen}(1994)}]{Rosen:1994}%
  \BibitemOpen
  \bibfield  {author} {\bibinfo {author} {\bibfnamefont {N}~\bibnamefont
  {Rosen}},\ }\bibfield  {title} {\enquote {\bibinfo {title} {A classical
  {P}roca particle},}\ }\href@noop {} {\bibfield  {journal} {\bibinfo
  {journal} {Found. Phys.}\ }\textbf {\bibinfo {volume} {24}},\ \bibinfo
  {pages} {1689--1695} (\bibinfo {year} {1994})}\BibitemShut {NoStop}%
\bibitem [{\citenamefont {Seitz}(1986)}]{Seitz:1986sc}%
  \BibitemOpen
  \bibfield  {author} {\bibinfo {author} {\bibfnamefont {M.}~\bibnamefont
  {Seitz}},\ }\bibfield  {title} {\enquote {\bibinfo {title} {Proca field in a
  space-time with curvature and torsion},}\ }\href {\doibase
  10.1088/0264-9381/3/6/023} {\bibfield  {journal} {\bibinfo  {journal} {Class.
  Quantum Grav.}\ }\textbf {\bibinfo {volume} {3}},\ \bibinfo {pages}
  {1265--1273} (\bibinfo {year} {1986})}\BibitemShut {NoStop}%
\bibitem [{\citenamefont {Goodsell}\ \emph {et~al.}(2009)\citenamefont
  {Goodsell}, \citenamefont {Jaeckel}, \citenamefont {Redondo},\ and\
  \citenamefont {Ringwald}}]{Goodsell:2009xc}%
  \BibitemOpen
  \bibfield  {author} {\bibinfo {author} {\bibfnamefont {Mark}\ \bibnamefont
  {Goodsell}}, \bibinfo {author} {\bibfnamefont {Joerg}\ \bibnamefont
  {Jaeckel}}, \bibinfo {author} {\bibfnamefont {Javier}\ \bibnamefont
  {Redondo}}, \ and\ \bibinfo {author} {\bibfnamefont {Andreas}\ \bibnamefont
  {Ringwald}},\ }\bibfield  {title} {\enquote {\bibinfo {title} {Naturally
  light hidden photons in {LARGE} volume string compactifications},}\ }\href
  {\doibase 10.1088/1126-6708/2009/11/027} {\bibfield  {journal} {\bibinfo
  {journal} {JHEP}\ }\textbf {\bibinfo {volume} {0911}},\ \bibinfo {pages}
  {027} (\bibinfo {year} {2009})},\ \Eprint {http://arxiv.org/abs/0909.0515}
  {arXiv:0909.0515 [hep-ph]} \BibitemShut {NoStop}%
\bibitem [{\citenamefont {Essig}\ \emph {et~al.}(2013)\citenamefont {Essig}
  \emph {et~al.}}]{Essig:2013lka}%
  \BibitemOpen
  \bibfield  {author} {\bibinfo {author} {\bibfnamefont {Rouven}\ \bibnamefont
  {Essig}} \emph {et~al.},\ }\bibfield  {title} {\enquote {\bibinfo {title}
  {Working group report: New light weakly coupled particles},}\ }in\ \href@noop
  {} {\emph {\bibinfo {booktitle} {{Proceedings, 2013 Community Summer Study on
  the Future of U.S. Particle Physics: Snowmass on the Mississippi (CSS2013):
  Minneapolis, MN, USA, July 29-August 6, 2013}}}}\ (\bibinfo {year} {2013})\
  \Eprint {http://arxiv.org/abs/1311.0029} {arXiv:1311.0029 [hep-ph]}
  \BibitemShut {NoStop}%
\bibitem [{\citenamefont {Arvanitaki}\ \emph {et~al.}(2010)\citenamefont
  {Arvanitaki}, \citenamefont {Dimopoulos}, \citenamefont {Dubovsky},
  \citenamefont {Kaloper},\ and\ \citenamefont
  {March-Russell}}]{Arvanitaki:2009fg}%
  \BibitemOpen
  \bibfield  {author} {\bibinfo {author} {\bibfnamefont {Asimina}\ \bibnamefont
  {Arvanitaki}}, \bibinfo {author} {\bibfnamefont {Savas}\ \bibnamefont
  {Dimopoulos}}, \bibinfo {author} {\bibfnamefont {Sergei}\ \bibnamefont
  {Dubovsky}}, \bibinfo {author} {\bibfnamefont {Nemanja}\ \bibnamefont
  {Kaloper}}, \ and\ \bibinfo {author} {\bibfnamefont {John}\ \bibnamefont
  {March-Russell}},\ }\bibfield  {title} {\enquote {\bibinfo {title} {String
  axiverse},}\ }\href {\doibase 10.1103/PhysRevD.81.123530} {\bibfield
  {journal} {\bibinfo  {journal} {Phys. Rev. D}\ }\textbf {\bibinfo {volume}
  {81}},\ \bibinfo {pages} {123530} (\bibinfo {year} {2010})},\ \Eprint
  {http://arxiv.org/abs/0905.4720} {arXiv:0905.4720 [hep-th]} \BibitemShut
  {NoStop}%
\bibitem [{\citenamefont {Baryakhtar}\ \emph {et~al.}(2017)\citenamefont
  {Baryakhtar}, \citenamefont {Lasenby},\ and\ \citenamefont
  {Teo}}]{Baryakhtar:2017ngi}%
  \BibitemOpen
  \bibfield  {author} {\bibinfo {author} {\bibfnamefont {Masha}\ \bibnamefont
  {Baryakhtar}}, \bibinfo {author} {\bibfnamefont {Robert}\ \bibnamefont
  {Lasenby}}, \ and\ \bibinfo {author} {\bibfnamefont {Mae}\ \bibnamefont
  {Teo}},\ }\bibfield  {title} {\enquote {\bibinfo {title} {Black hole
  superradiance signatures of ultralight vectors},}\ }\href {\doibase
  10.1103/PhysRevD.96.035019} {\bibfield  {journal} {\bibinfo  {journal} {Phys.
  Rev. D}\ }\textbf {\bibinfo {volume} {96}},\ \bibinfo {pages} {035019}
  (\bibinfo {year} {2017})},\ \Eprint {http://arxiv.org/abs/1704.05081}
  {arXiv:1704.05081 [hep-ph]} \BibitemShut {NoStop}%
\bibitem [{\citenamefont {Cardoso}\ \emph {et~al.}(2018)\citenamefont
  {Cardoso}, \citenamefont {Dias}, \citenamefont {Hartnett}, \citenamefont
  {Middleton}, \citenamefont {Pani},\ and\ \citenamefont
  {Santos}}]{Cardoso:2018tly}%
  \BibitemOpen
  \bibfield  {author} {\bibinfo {author} {\bibfnamefont {Vitor}\ \bibnamefont
  {Cardoso}}, \bibinfo {author} {\bibfnamefont {Oscar J.~C.}\ \bibnamefont
  {Dias}}, \bibinfo {author} {\bibfnamefont {Gavin~S.}\ \bibnamefont
  {Hartnett}}, \bibinfo {author} {\bibfnamefont {Matthew}\ \bibnamefont
  {Middleton}}, \bibinfo {author} {\bibfnamefont {Paolo}\ \bibnamefont {Pani}},
  \ and\ \bibinfo {author} {\bibfnamefont {Jorge~E.}\ \bibnamefont {Santos}},\
  }\bibfield  {title} {\enquote {\bibinfo {title} {Constraining the mass of
  dark photons and axion-like particles through black-hole superradiance},}\
  }\href@noop {} {\bibfield  {journal} {\bibinfo  {journal} {JCAP}\ }\textbf
  {\bibinfo {volume} {1803}},\ \bibinfo {pages} {043} (\bibinfo {year}
  {2018})},\ \bibinfo {note} {preprint},\ \Eprint
  {http://arxiv.org/abs/1801.01420} {arXiv:1801.01420 [gr-qc]} \BibitemShut
  {NoStop}%
\bibitem [{\citenamefont {Pani}\ \emph
  {et~al.}(2012{\natexlab{a}})\citenamefont {Pani}, \citenamefont {Cardoso},
  \citenamefont {Gualtieri}, \citenamefont {Berti},\ and\ \citenamefont
  {Ishibashi}}]{Pani:2012vp}%
  \BibitemOpen
  \bibfield  {author} {\bibinfo {author} {\bibfnamefont {Paolo}\ \bibnamefont
  {Pani}}, \bibinfo {author} {\bibfnamefont {Vitor}\ \bibnamefont {Cardoso}},
  \bibinfo {author} {\bibfnamefont {Leonardo}\ \bibnamefont {Gualtieri}},
  \bibinfo {author} {\bibfnamefont {Emanuele}\ \bibnamefont {Berti}}, \ and\
  \bibinfo {author} {\bibfnamefont {Akihiro}\ \bibnamefont {Ishibashi}},\
  }\bibfield  {title} {\enquote {\bibinfo {title} {Black hole bombs and photon
  mass bounds},}\ }\href {\doibase 10.1103/PhysRevLett.109.131102} {\bibfield
  {journal} {\bibinfo  {journal} {Phys. Rev. Lett.}\ }\textbf {\bibinfo
  {volume} {109}},\ \bibinfo {pages} {131102} (\bibinfo {year}
  {2012}{\natexlab{a}})},\ \Eprint {http://arxiv.org/abs/1209.0465}
  {arXiv:1209.0465 [gr-qc]} \BibitemShut {NoStop}%
\bibitem [{\citenamefont {Rosa}\ and\ \citenamefont
  {Dolan}(2012)}]{Rosa:2011my}%
  \BibitemOpen
  \bibfield  {author} {\bibinfo {author} {\bibfnamefont {Joao~G.}\ \bibnamefont
  {Rosa}}\ and\ \bibinfo {author} {\bibfnamefont {Sam~R.}\ \bibnamefont
  {Dolan}},\ }\bibfield  {title} {\enquote {\bibinfo {title} {Massive vector
  fields on the {S}chwarzschild spacetime: quasi-normal modes and bound
  states},}\ }\href {\doibase 10.1103/PhysRevD.85.044043} {\bibfield  {journal}
  {\bibinfo  {journal} {Phys. Rev. D}\ }\textbf {\bibinfo {volume} {85}},\
  \bibinfo {pages} {044043} (\bibinfo {year} {2012})},\ \Eprint
  {http://arxiv.org/abs/1110.4494} {arXiv:1110.4494 [hep-th]} \BibitemShut
  {NoStop}%
\bibitem [{\citenamefont {Pani}\ \emph
  {et~al.}(2012{\natexlab{b}})\citenamefont {Pani}, \citenamefont {Cardoso},
  \citenamefont {Gualtieri}, \citenamefont {Berti},\ and\ \citenamefont
  {Ishibashi}}]{Pani:2012bp}%
  \BibitemOpen
  \bibfield  {author} {\bibinfo {author} {\bibfnamefont {Paolo}\ \bibnamefont
  {Pani}}, \bibinfo {author} {\bibfnamefont {Vitor}\ \bibnamefont {Cardoso}},
  \bibinfo {author} {\bibfnamefont {Leonardo}\ \bibnamefont {Gualtieri}},
  \bibinfo {author} {\bibfnamefont {Emanuele}\ \bibnamefont {Berti}}, \ and\
  \bibinfo {author} {\bibfnamefont {Akihiro}\ \bibnamefont {Ishibashi}},\
  }\bibfield  {title} {\enquote {\bibinfo {title} {Perturbations of slowly
  rotating black holes: massive vector fields in the {K}err metric},}\ }\href
  {\doibase 10.1103/PhysRevD.86.104017} {\bibfield  {journal} {\bibinfo
  {journal} {Phys. Rev. D}\ }\textbf {\bibinfo {volume} {86}},\ \bibinfo
  {pages} {104017} (\bibinfo {year} {2012}{\natexlab{b}})},\ \Eprint
  {http://arxiv.org/abs/1209.0773} {arXiv:1209.0773 [gr-qc]} \BibitemShut
  {NoStop}%
\bibitem [{\citenamefont {Endlich}\ and\ \citenamefont
  {Penco}(2017)}]{Endlich:2016jgc}%
  \BibitemOpen
  \bibfield  {author} {\bibinfo {author} {\bibfnamefont {Solomon}\ \bibnamefont
  {Endlich}}\ and\ \bibinfo {author} {\bibfnamefont {Riccardo}\ \bibnamefont
  {Penco}},\ }\bibfield  {title} {\enquote {\bibinfo {title} {A modern approach
  to superradiance},}\ }\href {\doibase 10.1007/JHEP05(2017)052} {\bibfield
  {journal} {\bibinfo  {journal} {JHEP}\ }\textbf {\bibinfo {volume} {1705}},\
  \bibinfo {pages} {052} (\bibinfo {year} {2017})},\ \Eprint
  {http://arxiv.org/abs/1609.06723} {arXiv:1609.06723 [hep-th]} \BibitemShut
  {NoStop}%
\bibitem [{\citenamefont {East}\ and\ \citenamefont
  {Pretorius}(2017)}]{East:2017ovw}%
  \BibitemOpen
  \bibfield  {author} {\bibinfo {author} {\bibfnamefont {William~E.}\
  \bibnamefont {East}}\ and\ \bibinfo {author} {\bibfnamefont {Frans}\
  \bibnamefont {Pretorius}},\ }\bibfield  {title} {\enquote {\bibinfo {title}
  {Superradiant instability and backreaction of massive vector fields around
  kerr black holes},}\ }\href {\doibase 10.1103/PhysRevLett.119.041101}
  {\bibfield  {journal} {\bibinfo  {journal} {Phys. Rev. Lett.}\ }\textbf
  {\bibinfo {volume} {119}},\ \bibinfo {pages} {041101} (\bibinfo {year}
  {2017})},\ \Eprint {http://arxiv.org/abs/1704.04791} {arXiv:1704.04791
  [gr-qc]} \BibitemShut {NoStop}%
\bibitem [{\citenamefont {East}(2017)}]{East:2017mrj}%
  \BibitemOpen
  \bibfield  {author} {\bibinfo {author} {\bibfnamefont {William~E.}\
  \bibnamefont {East}},\ }\bibfield  {title} {\enquote {\bibinfo {title}
  {Superradiant instability of massive vector fields around spinning black
  holes in the relativistic regime},}\ }\href {\doibase
  10.1103/PhysRevD.96.024004} {\bibfield  {journal} {\bibinfo  {journal} {Phys.
  Rev. D}\ }\textbf {\bibinfo {volume} {96}},\ \bibinfo {pages} {024004}
  (\bibinfo {year} {2017})},\ \Eprint {http://arxiv.org/abs/1705.01544}
  {arXiv:1705.01544 [gr-qc]} \BibitemShut {NoStop}%
\bibitem [{\citenamefont {Frolov}\ \emph {et~al.}(2017)\citenamefont {Frolov},
  \citenamefont {Krtou\v{s}},\ and\ \citenamefont
  {Kubiz\v{n}\'ak}}]{FrolovKrtousKubiznak:2017review}%
  \BibitemOpen
  \bibfield  {author} {\bibinfo {author} {\bibfnamefont {Valeri~P.}\
  \bibnamefont {Frolov}}, \bibinfo {author} {\bibfnamefont {Pavel}\
  \bibnamefont {Krtou\v{s}}}, \ and\ \bibinfo {author} {\bibfnamefont {David}\
  \bibnamefont {Kubiz\v{n}\'ak}},\ }\bibfield  {title} {\enquote {\bibinfo
  {title} {Black holes, hidden symmetries, and complete integrability},}\
  }\href {\doibase 10.1007/s41114-017-0009-9} {\bibfield  {journal} {\bibinfo
  {journal} {Living Rev. Rel.}\ }\textbf {\bibinfo {volume} {20}},\ \bibinfo
  {pages} {6} (\bibinfo {year} {2017})},\ \Eprint
  {http://arxiv.org/abs/1705.05482} {arXiv:1705.05482 [gr-qc]} \BibitemShut
  {NoStop}%
\bibitem [{\citenamefont {Chen}\ \emph {et~al.}(2006)\citenamefont {Chen},
  \citenamefont {Lu},\ and\ \citenamefont {Pope}}]{Chen:2006xh}%
  \BibitemOpen
  \bibfield  {author} {\bibinfo {author} {\bibfnamefont {W.}~\bibnamefont
  {Chen}}, \bibinfo {author} {\bibfnamefont {H.}~\bibnamefont {Lu}}, \ and\
  \bibinfo {author} {\bibfnamefont {C.~N.}\ \bibnamefont {Pope}},\ }\bibfield
  {title} {\enquote {\bibinfo {title} {General {K}err-{NUT}-{AdS} metrics in
  all dimensions},}\ }\href {\doibase 10.1088/0264-9381/23/17/013} {\bibfield
  {journal} {\bibinfo  {journal} {Class. Quantum Grav.}\ }\textbf {\bibinfo
  {volume} {23}},\ \bibinfo {pages} {5323--5340} (\bibinfo {year} {2006})},\
  \Eprint {http://arxiv.org/abs/hep-th/0604125} {arXiv:hep-th/0604125}
  \BibitemShut {NoStop}%
\bibitem [{\citenamefont {Lunin}(2017)}]{Lunin:2017}%
  \BibitemOpen
  \bibfield  {author} {\bibinfo {author} {\bibfnamefont {Oleg}\ \bibnamefont
  {Lunin}},\ }\bibfield  {title} {\enquote {\bibinfo {title} {Maxwell's
  equations in the {Myers-Perry} geometry},}\ }\href {\doibase
  10.1007/JHEP12(2017)138} {\bibfield  {journal} {\bibinfo  {journal} {JHEP}\
  }\textbf {\bibinfo {volume} {1712}},\ \bibinfo {pages} {138} (\bibinfo {year}
  {2017})},\ \Eprint {http://arxiv.org/abs/1708.06766} {arXiv:1708.06766
  [hep-th]} \BibitemShut {NoStop}%
\bibitem [{\citenamefont {Frolov}\ \emph {et~al.}(2018)\citenamefont {Frolov},
  \citenamefont {Krtou\v{s}},\ and\ \citenamefont
  {Kubiz\v{n}\'ak}}]{FrolovKrtousKubiznak:2018a}%
  \BibitemOpen
  \bibfield  {author} {\bibinfo {author} {\bibfnamefont {Valeri~P.}\
  \bibnamefont {Frolov}}, \bibinfo {author} {\bibfnamefont {Pavel}\
  \bibnamefont {Krtou\v{s}}}, \ and\ \bibinfo {author} {\bibfnamefont {David}\
  \bibnamefont {Kubiz\v{n}\'ak}},\ }\bibfield  {title} {\enquote {\bibinfo
  {title} {Separation variables in {M}axwell equations in
  {Pleba\'{n}ski--Demia\'{n}ski metric}},}\ }\href {\doibase
  10.1103/PhysRevD.97.101701} {\bibfield  {journal} {\bibinfo  {journal} {Phys.
  Rev. D}\ }\textbf {\bibinfo {volume} {97}},\ \bibinfo {pages} {101701(R)}
  (\bibinfo {year} {2018})},\ \Eprint {http://arxiv.org/abs/1802.09491}
  {arXiv:1802.09491 [hep-th]} \BibitemShut {NoStop}%
\bibitem [{\citenamefont {Krtou\v{s}}\ \emph {et~al.}(2018)\citenamefont
  {Krtou\v{s}}, \citenamefont {Frolov},\ and\ \citenamefont
  {Kubiz\v{n}\'ak}}]{KrtousEtal:2018}%
  \BibitemOpen
  \bibfield  {author} {\bibinfo {author} {\bibfnamefont {Pavel}\ \bibnamefont
  {Krtou\v{s}}}, \bibinfo {author} {\bibfnamefont {Valeri~P.}\ \bibnamefont
  {Frolov}}, \ and\ \bibinfo {author} {\bibfnamefont {David}\ \bibnamefont
  {Kubiz\v{n}\'ak}},\ }\bibfield  {title} {\enquote {\bibinfo {title}
  {Separation of {M}axwell equations in {Kerr--NUT--(A)dS} spacetimes},}\
  }\href@noop {} {\  (\bibinfo {year} {2018})},\ \bibinfo {note} {preprint},\
  \Eprint {http://arxiv.org/abs/1803.02485} {arXiv:1803.02485 [hep-th]}
  \BibitemShut {NoStop}%
\bibitem [{\citenamefont {Konoplya}\ \emph {et~al.}(2007)\citenamefont
  {Konoplya}, \citenamefont {Zhidenko},\ and\ \citenamefont
  {Molina}}]{Konoplya:2006gq}%
  \BibitemOpen
  \bibfield  {author} {\bibinfo {author} {\bibfnamefont {R.~A.}\ \bibnamefont
  {Konoplya}}, \bibinfo {author} {\bibfnamefont {A.}~\bibnamefont {Zhidenko}},
  \ and\ \bibinfo {author} {\bibfnamefont {C.}~\bibnamefont {Molina}},\
  }\bibfield  {title} {\enquote {\bibinfo {title} {Late time tails of the
  massive vector field in a black hole background},}\ }\href {\doibase
  10.1103/PhysRevD.75.084004} {\bibfield  {journal} {\bibinfo  {journal} {Phys.
  Rev. D}\ }\textbf {\bibinfo {volume} {75}},\ \bibinfo {pages} {084004}
  (\bibinfo {year} {2007})},\ \Eprint {http://arxiv.org/abs/gr-qc/0602047}
  {arXiv:gr-qc/0602047} \BibitemShut {NoStop}%
\bibitem [{\citenamefont {Konoplya}(2006)}]{Konoplya:2005hr}%
  \BibitemOpen
  \bibfield  {author} {\bibinfo {author} {\bibfnamefont {R.~A.}\ \bibnamefont
  {Konoplya}},\ }\bibfield  {title} {\enquote {\bibinfo {title} {Massive vector
  field perturbations in the {S}chwarzschild background: {S}tability and
  unusual quasinormal spectrum},}\ }\href {\doibase 10.1103/PhysRevD.73.024009}
  {\bibfield  {journal} {\bibinfo  {journal} {Phys. Rev. D}\ }\textbf {\bibinfo
  {volume} {73}},\ \bibinfo {pages} {024009} (\bibinfo {year} {2006})},\
  \Eprint {http://arxiv.org/abs/gr-qc/0509026} {arXiv:gr-qc/0509026}
  \BibitemShut {NoStop}%
\bibitem [{\citenamefont
  {Shlapentokh-Rothman}(2014)}]{Shlapentokh-Rothman:2013ysa}%
  \BibitemOpen
  \bibfield  {author} {\bibinfo {author} {\bibfnamefont {Yakov}\ \bibnamefont
  {Shlapentokh-Rothman}},\ }\bibfield  {title} {\enquote {\bibinfo {title}
  {Exponentially growing finite energy solutions for the {K}lein-{G}ordon
  equation on sub-extremal {K}err spacetimes},}\ }\href {\doibase
  10.1007/s00220-014-2033-x} {\bibfield  {journal} {\bibinfo  {journal}
  {Commun. Math. Phys.}\ }\textbf {\bibinfo {volume} {329}},\ \bibinfo {pages}
  {859--891} (\bibinfo {year} {2014})},\ \Eprint
  {http://arxiv.org/abs/1302.3448} {arXiv:1302.3448 [gr-qc]} \BibitemShut
  {NoStop}%
\bibitem [{\citenamefont {Detweiler}(1980)}]{Detweiler:1980uk}%
  \BibitemOpen
  \bibfield  {author} {\bibinfo {author} {\bibfnamefont {Steven~L.}\
  \bibnamefont {Detweiler}},\ }\bibfield  {title} {\enquote {\bibinfo {title}
  {{Klein-Gordon} equation and rotating black holes},}\ }\href {\doibase
  10.1103/PhysRevD.22.2323} {\bibfield  {journal} {\bibinfo  {journal} {Phys.
  Rev. D}\ }\textbf {\bibinfo {volume} {22}},\ \bibinfo {pages} {2323--2326}
  (\bibinfo {year} {1980})}\BibitemShut {NoStop}%
\end{thebibliography}

%

\end{document}